%% file: conference_101719.tex
\documentclass[conference]{IEEEtran}
\IEEEoverridecommandlockouts
\usepackage{cite}
\usepackage{amsmath,amssymb,amsfonts}
\usepackage{algorithmic}
\usepackage{graphicx}
\usepackage{textcomp}
\usepackage{xcolor}
\usepackage{float}
\usepackage{booktabs}
\usepackage{todonotes}
\usepackage{subcaption}

\def\BibTeX{{\rm B\kern-.05em{\sc i\kern-.025em b}\kern-.08em
    T\kern-.1667em\lower.7ex\hbox{E}\kern-.125emX}}
\begin{document}

\title{M3Net: A Multi-Metric Mixture of Experts Network Digital Twin with Graph Neural Networks
}

\author{
\IEEEauthorblockN{Blessed Guda}
\IEEEauthorblockA{\textit{Carnegie Mellon University} \\
\textit{Kigali, Rwanda}\\
blessedg@andrew.cmu.edu}
\and

\IEEEauthorblockN{Carlee Joe-Wong}
\IEEEauthorblockA{\textit{Carnegie Mellon University} \\
\textit{Pittsburgh, USA}\\
cjoewong@andrew.cmu.edu}

}

\maketitle

\input{sections/abstract}


\input{sections/intro}

\input{sections/related}

\input{sections/formulation}
\input{sections/architecture}

\input{sections/implementation}

\input{sections/dataset}

\input{sections/evaluation}

\section{Conclusion}\label{sec:conclusion}
In this paper, we have presented several advancements in network modeling using Graph Neural Networks, specifically focusing on improving the RouteNet-Fermi architecture. 
Our results demonstrate that incorporating node features and a gating mechanism enhances the accuracy of delay predictions. The hierarchical model for jitter and packet loss estimation showed substantial improvements, marking a significant step forward in network modeling using real-world data.
The M3Net architecture, particularly with 4 experts, proved to be more effective than the baseline model and the version with 8 experts, highlighting the importance of finding the right balance in model complexity.
Furthermore, our efficient implementation techniques resulted in significant speed-ups, enabling better utilization of GPU resources and allowing for more diverse training samples per batch.
These advancements contribute to more accurate and scalable network digital twins, which are crucial for planning and managing complex modern networks. Future work could explore the application of these techniques to larger, more diverse network datasets and investigate their effectiveness in real-time network optimization scenarios, e.g., network slicing or flow admission control.


\bibliographystyle{IEEEtran}
\bibliography{references}

\end{document}

%% file: sections/abstract.tex
\begin{abstract}
    The rise of 5G/6G network technologies promises to enable applications like autonomous vehicles and virtual reality, resulting in a significant increase in connected devices and necessarily complicating network management. Even worse, these applications often have strict, yet heterogeneous, performance requirements across metrics like latency and reliability. Much recent work has thus focused on developing the ability to predict network performance. However, traditional methods for network modeling, like discrete event simulators and emulation, often fail to balance accuracy and scalability. Network Digital Twins (NDTs), augmented by machine learning, present a viable solution by creating virtual replicas of physical networks for real-time simulation and analysis.
    State-of-the-art models, however, fall short of full-fledged NDTs, as they often focus only on a single performance metric or simulated network data. 
    We introduce M3Net, a Multi-Metric Mixture-of-experts (MoE) NDT that uses a graph neural network architecture to estimate multiple performance metrics from an expanded set of network state data in a range of scenarios. 
    We show that M3Net significantly enhances the accuracy of flow delay predictions by reducing the MAPE (Mean Absolute Percentage Error) from 20.06\% to 17.39\%, while also achieving 66.47\% and 78.7\% accuracy on jitter and packets dropped for each flow. 
\end{abstract}

%% file: sections/intro.tex
\section{Introduction}

Emerging 5G and 6G mobile network architectures aim to support new applications like autonomous vehicles and mixed reality~\cite{ericsson-6g,almasan2022digital}, both of which require significantly expanded network capabilities. These and other new applications envisioned as part of the 5G and 6G network ecosystem will lead to massive numbers of connected devices with heterogeneous performance expectations, which increases the complexity and cost of managing communication networks~\cite{almasan2022digital}. For example, interactive applications like augmented reality generally require response latencies under 200ms~\cite{braud2017future}, while safety-critical applications like autonomous vehicles might require highly reliable delivery of high-priority packets~\cite{dong2022use}. Planning for and meeting these diverse requirements is difficult, due to the increasing complexity of network architectures and number of devices they must support. Direct experimentation on a production network can lead to service disruptions for users~\cite{suarez2015computer,tran1992performance}, making network modeling a crucial tool for designing and managing networks\cite{lu2013particle}. Such models, for example, can be used to predict whether a particular user's application can be supported under current network conditions or to predict anomalies in performance that require operator intervention. 

Developing accurate network models is a long-standing challenge in the networking community~\cite{suarez2015computer,tran1992performance}, which has only gotten worse in today's mobile networks. Experimenting with real network equipment is expensive and time consuming, especially since it is infeasible to build a full physical prototype of a national or global-scale network. These limitations mean that network modeling is often achieved using discrete event network simulators such as Omnet++~\cite{omnet} and network simulator (ns-3)~\cite{ns}, emulation~\cite{lantz2010network}, or using analytical models like queueing theory \cite{tran1992performance}. However, network simulators often model the network at a packet level, which does not scale to large numbers of devices and large amounts of network traffic~\cite{li2024splitsim} as the simulation time scales linearly with the number of packets. Meanwhile, network emulators prioritize speed, but often compromise accuracy, rendering them impractical. In this work, we develop a model that abstracts away from packet-level logic to predict \textit{flow-level performance metrics}, with the aim of accurately characterizing average performance (e.g., delay, packets dropped) for a given flow, given its surrounding network environment~\cite{suarez2021graph,ferriol2023routenet}.

\subsection{Research Challenges for Network Digital Twins}

Network Digital Twins (NDTs), which create virtual replicas of physical networks, have recently been proposed to address the scalability challenges of traditional modeling techniques. 
Machine learning-based NDT approaches in particular can mitigate \textbf{scalability challenges} by learning patterns from a large amount of data, thereby eliminating the need for packet-by-packet simulation while maintaining high accuracy \cite{mestres2018understanding,carner2017machine}. Relying on machine learning, however, introduces its own scalability challenges. First, \textit{networks are complex environments}, with many features that can affect network performance, e.g., the network topology, the path of the flow, and the state of any competing flows that share its links. Thus, the model must have sufficient capacity to capture these complexities and use them to accurately predict different performance measures. Second, NDTs should be able to  \textit{make accurate predictions for a variety of network environments}: for example, as the number of competing flows on a link varies, each flow's performance may vary; our model should be flexible enough to recognize and adapt to these changes~\cite{hui2022digital}. 

Finally, \textit{training these models can be inefficient}. Typically, large-scale machine learning uses GPU machines to accelerate the training.
In the case we consider,  different network configurations contain different sets of links and devices used. For each network configuration, a variable number of flows is sent through the network. Also, each of the flows goes through paths of varying lengths. 
These unique features of NDTs precludes a straightforward approach to control and increase the batch sizes of the data to take advantage of speedups offered by GPUs. All implementations we have studied are limited to batch sizes pre-determined by the batch flow sizes sent for each configuration during the data generation process. 


Recent advances in networking have led to the development of powerful \textbf{Graph Neural Networks (GNNs) that can mimic complex network environments} effectively~\cite{rusek2020routenet,ferriol2023routenet,ferriol2022building} and thus serve as a promising basis for a NDT, solving the first challenge of modeling complex network environments. Indeed, using a GNN is a natural choice for computer networks due to the graph structure of network data. While most GNN models have only been been validated with data generated from simulators, recent initiatives have included evaluations on more realistic data from network testbeds. State-of-the-art GNN models~\cite{ferriol2023routenet}, for example, were recently developed for the 2023 edition of the Graph Neural Networking challenge~\cite{suarez2021graph}, which tasked participants with building a GNN-based Network Digital Twin using real network data generated from a small testbed. The challenge involves modeling traffic, topology, and configuration to accurately estimate network performance, specifically focusing on delay. Participants were required to predict the mean delay for each flow based on network topology, flow packet traces, and routing configuration.

While these recent initiatives have led to several GNN architectures that can predict network delay, \textbf{these models fall short of NDTs and do not fully address the challenges outlined above of learning-based NDTs}. First, while they can capture some of the complexity of network environments, these architectures focus on modeling the network links but do not explicitly model the state of the network nodes (routers and switches). We solve this limitation by adding extra node features to make the network aware of the devices. Second, we find that, while such architectures (including the winner of the 2023 GNN Challenge) have low average delays, they often struggle with correct predictions in specific network scenarios. Even worse, they cannot predict other important metrics like jitter and packets dropped; we find that even the best-performing GNN models in terms of average delay predictions do not directly translate to estimate jitter and packet loss~\cite{ferriol2023routenet}. We address these challenges by proposing "\textit{\textbf{M3Net}}" (\textbf{M}ulti-\textbf{M}etric \textbf{M}ixture of experts \textbf{Net}work digital twin), a gating approach inspired by a Mixture of Experts \cite{shazeer2017outrageously} to have different experts focus on different categories of network flows. We show that this gated architecture is particularly effective for predicting jitter and packet loss. Finally, current models are difficult to implement on GPUs; their open-source implementations are primarily CPU-based, significantly slowing the training time.

\subsection{Our Contributions}

The \textbf{key contributions} and findings of this research are summarized as follows:
\begin{itemize}
    
    \item We find out that na\"ively using the state-of-the-art Routenet fermi architecture~\cite{ferriol2023routenet} produces inaccurate predictions of jitter and packet loss. Thus, we propose a \textbf{new GNN architecture, M3Net,} that (i) explicitly predicts not just packet delay but also jitter and packet loss, and (ii) uses a hierarchical structure to ensure accurate predictions in a range of network environments.
    \item We introduce an \textbf{efficient GPU-based implementation} of M3Net, which allows for significantly faster training than prior CPU-based NDTs. We provide an open-source release of our implementation of M3Net.\footnote{Link redacted to preserve double-blind submission anonymity.}
    \item We evaluate our M3Net architecture on a real network dataset to \textbf{explore the relationships} between the delay, jitter, and packet loss performance metrics in different scenarios. 
    We surprisingly find that, unlike what we would intuitively expect, there is no correlation between the amount of packets generated, average traffic bandwidth, and the network delay.
    \item We \textbf{evaluate our M3Net architecture} on both Multi Burst (MB) only traffic and flows containing both Constant Bit Rate (CBR) and MB traffic. We show that M3Net accurately predicts both jitter and packets dropped and achieves 17.8\% mean absolute percentage error (MAPE) in delay prediction on flows containing both Constant Bit Rate (CBR) and MB traffic. This is a significant improvement to an MLP (multi-layer perceptron) with MAPE of 56.10\%, the Routenet-F  with 43.38\%,  and a Routenet-F with attention with 20.06\%  
\end{itemize}

In Section~\ref{sec:related}, we give an overview of related work before formulating our problem of predicting delay, jitter, and packet loss in Section~\ref{sec:formulation}. We then introduce M3Net in Section~\ref{sec:architecture}, as well as our efficient implementation in Section~\ref{sec:implementation}. We then describe our evaluation dataset (Section~\ref{sec:dataset}) and results (Section~\ref{sec:evaluation}) before concluding in Section~\ref{sec:conclusion}.

%% file: sections/related.tex
\section{Related Work}\label{sec:related}

Many early versions of machine learning-based NDTs utilized multi-layer perceptron (MLP) and recurrent neural network (RNN) architectures, which are not designed to capture topological information in the network~\cite{hui2022digital,mestres2018understanding}. Unlike our work, these early network models also generally focused on predicting a single performance metric of interest~\cite{hui2022digital}, instead of combining models to build a true digital twin that can predict multiple performance metrics.

\textbf{Graph neural networks}, which can naturally encode topological information that may be useful for incorporating information on a flow's path through the network, are a natural choice for making flow-level performance predictions. One early example of such a GNN is TwinNet~\cite{ferriol2022building}, which takes as input the states of the links along a flow's path as well as the queues on each of these links. TwinNet then uses cylic message passing with gated recurrent units (GRUs) to construct a path embedding, with a two-layer MLP (multi-layer perceptron) that converts this embedding to the final predicted performance. However, TwinNet does not generalize well to unseen network routing configurations, or to longer paths and larger link capacities.

In an attempt to scale better to longer paths with larger link capacities, \cite{ferriol2023routenet} proposed Route Net-Fermi, an improvement of the original RouteNet architecture~\cite{rusek2020routenet}. Instead of predicting the mean-per-flow delay directly from the final hidden path hidden state (unlike in TwinNet and RouteNet-Erlang~\cite{ferriol2022routenet}), it is accumulated from the hidden states of the links that the flow traverses, and the delay is predicted by dividing the predicted queue occupancy by the link capacity. 
This construction helps in scaling to longer path lengths and higher link capacities, as the readout is independent of the link capacities and flow lengths, which along with incorporating extra input features, allows for more accurate average delay predictions. 
\cite{fu2023active} provided an active and few shot-learning framework which allows the RouteNet-Fermi model to learn from a constrained number of data samples and nodes. 
Unlike these works, we use GNNs to predict not just average delay but also \textit{jitter} and the \textit{number of dropped packets}. Moreover, we use a novel \textit{mixture-of-experts} (MoE) gating mechanism to handle predictions in a variety of network scenarios, which we show significantly improves prediction performance.

Several newer approaches have attempted to incorporate the power of the \textbf{Graph Attention (GAT) mechanism} \cite{velivckovic2017graph}, which typically enhances GNNs' predictive power, into estimation models. Most recently, \cite{modesto2023graph} computed attention weights across each dimension of the hidden states in the RouteNet fermi model, while 
\cite{dhamala2024performance}
incorporated the attention mechanism into the path update step of the RouteNet model\cite{rusek2020routenet}. However, the attention mechanisms used in these models are completely local to the internal dimensions of the path states, even though intuitively attention should consider neighbourhood information of other links in the network. 

\textbf{Applications of NDTs within networks} have also been a recent focus in the literature, including for network slicing~\cite{wang2020graph}, task offloading in edge computing~\cite{bellavista2021application,liu2021digital}, and network planning~\cite{zhao2022design}. Other popular NDTs focus on wireless applications and developing digital twins of radio access networks~\cite{vila2023design,khan2022digital} instead of path-based performance predictions. Unlike these works we focus primarily on flow-based performance predictions, which can then be used for applications such as network slicing or task offloading.

%% file: sections/formulation.tex
\section{Background and Problem Formulation}\label{sec:formulation}

We consider a computer network as a system of interconnected devices that communicate and share resources together. The network can be represented as a graph of $N$ nodes, representing network \textit{devices} (such as routers and switches), and $E$ edges, which represent \textit{links} that connect these devices~\cite{peterson2007computer}. When data within a \textit{flow} is to be transmitted from a source to destination, it follows a \textit{path} determined by a routing algorithm~\cite{duato2003interconnection}. The path represent sequences of links and nodes through which data travels from a source to destination.

\textbf{Performance prediction task.} We suppose that a sequence of flows $f = 1,2,\ldots$ enters the network, with each flow assigned to a given path $p(f)$ by a given routing algorithm. Our task is to predict the performance of each flow, which we denote by $y$. This performance may  be the average delay, packets dropped, or jitter of the flow. We choose these three metrics as they are commonly considered in prior work~\cite{ferriol2023routenet} and significantly affect the utility of flows for many envisioned 5G and 6G applications like mixed reality and autonomous driving \cite{platt2003adaptive}. Our objective is to solve

\begin{equation}\label{eq:1}
    \min_\theta \sum_{f = 1}^F \ell\left(y, g_\theta\left(\mathbf{x}_f, \left\{\mathbf{x}_{l\in p(f)}\right\}\right),\left\{\mathbf{x}_{l\not\in p(f)}\right\}\right),
\end{equation}
where $f = 1,2,\ldots,F$ indexes the flows in our training data, $\ell$ is a given loss function, e.g., mean-squared error, and $g$ represents our predictive model, which has learnable parameters $\theta$. We use $\mathbf{x}_f$ to denote the input features corresponding to flow $f$, and $\mathbf{x}_{l\in p(f)}$ to denote the input features for each link $l$ along the path $p(f)$. Note that our inputs also include links not part of $p(f)$; as we explain below in introducing our network states, these features can capture characteristics of competing flows $f'$ whose paths may share some links with $f$, which therefore may influence flow $f$'s performance. Since $y$ may represent one of three different flow performance metrics, we use a different parameterized model $g_\theta$ to estimate each metric. 

\textbf{Input features.} The elements of the features we used for the flows and links are given in Table \ref{tab:flow_link_metrics}. 
These are designed to capture aspects of the network environment that may affect flow performance, such as varying levels of congestion or processing loads. For instance, during peak usage times, a router might become a bottleneck due to high traffic volumes, leading to increased latency and packet loss. In our feature vector, this would be reflected in changes in the \textit{link load}. 
Similarly, the links may demonstrate fluctuating capacities and latencies based on the volume and priority of the traffic they carry. For example, a link might support high-speed data transfer under normal conditions but could experience reduced throughput if it becomes congested or if there is a high-priority traffic flow that preempts other data. These effects would be captured in our \textit{link capacity} feature. Additionally, external factors such as physical damage to cables or interference in wireless connections can further impact link performance.

Unlike prior work, we include the \textit{node degree} of each the device that the link is connected to as one of the link features. Intuitively, this feature crucially reflects the resource sharing that may be happening at each device due to other links connected to the same device. We find the simple addition of this feature to be important in increasing our model's prediction performance as well. 



\begin{table}[ht!]
\centering
\caption{Input flow and link features to our model.}
\begin{tabular}{|l|l|}
\hline
\textbf{Flow Features} & \textbf{Link} \\ \hline
Flow traffic & Link Capacity \\ \hline
Flow packets per burst & Link Load \\ \hline
Flow packet size & Normalized Link Load (per experiment) \\ \hline
Flow 90th percentile & Node degree of connected device \\ \hline
Flow interpacket mean &  \\ \hline
Flow interpacket variance &  \\ \hline
\end{tabular}

\label{tab:flow_link_metrics}
\end{table}
 
\textbf{Network link and path states.} As computer networks are dynamic systems \cite{pandey2019comprehensive}, our model converts the input link features $\mathbf{x}_l$ into \textit{states} for each link and path, which may depend on current network traffic and other network conditions. These states are then used to predict flow performance.
We use $\mathbf{h}_l$ to denote the state of each link $l$, which are initialized from the link features $\mathbf{x}_l$, and we use $\mathbf{h}_p$ to denote the state of each path $p = 1,2,\ldots,P$ which are derived from the input flow features $\mathbf{x}_f$ and corresponding link states $\mathbf{h}_l$ along path $p$. 

The network modeling approach we follow is designed to capture the cyclic dependency that exists between paths and links \cite{ferriol2022building}. We exclude queue states in this real network data scenario, as the states of queues may not be known in practice. This cyclic dependency is defined as follows:
\begin{itemize}
    \item The links along a path determine the \textbf{state of the path}.
     For a path $p$ of length n,
    \begin{equation}
         \mathbf{h}_{p} = F_{p}\left( \mathbf{h}_{l}^{1}, \mathbf{h}_{l}^{2}, ..., \mathbf{h}_{l}^{n}\right) 
    \end{equation}
    where $\mathbf{h}_{l}^{i}$ is the state of link $i$ in path $p$.
   
    \item The \textbf{state of a link} is influenced by the states of all paths that transmit traffic over it.
    For a link $l$ with $n$ such paths, 
    \begin{equation}
        \mathbf{h}_{l} = F_{l}( \mathbf{h}_{p}^{1}, \mathbf{h}_{p}^{2}, ..., \mathbf{h}_{p}^{n} ) 
    \end{equation}
    where $\mathbf{h}_{p}^{j}$ is the state of the $j$th path that includes link $l$.
\end{itemize}
Therefore, the learning problem is to learn approximations of the embedding functions $F_{p}$ and $F_{l}$ using $g_\theta$; we can then use the path and link state vectors, as well as the flow features, to predict flow performance. We describe the GNN architecture with which we do so in the next section.

%% file: sections/architecture.tex
\section{Our Architecture} \label{sec:architecture}

In this section, we introduce the structure of our GNN model for predicting flow performance. We first introduce the basic structure of our model, which is based on state-of-the-art GNN model RouteNet-Fermi~\cite{ferriol2023routenet}, in Section~\ref{sec:architecture-fermi}. We then discuss in detail the two main additions we have made to this architecture, which allows it to predict multiple performance metrics (Section~\ref{sec:architecture-multi}) and generalize better to different network states (Section~\ref{sec:architecture-gated}). M3Net's overall architecture is shown in Figure~\ref{fig:gatedroutenetarchiecture}.

\subsection{RouteNet Fermi GNN Architecture}\label{sec:architecture-fermi}

The learning procedure of the RouteNet-Fermi has three stages, namely the (i) initialization, (ii) message passing, and (iii) read-out phases. We follow this approach, as detailed in Figure~\ref{fig:gatedroutenetarchiecture}.

In the \textbf{initialization} phase, the states of the links and paths are initialized using embeddings of their feature values. 
Mathematically, RouteNet-Fermi computes
    \begin{equation*}
        \mathbf{h}_l^{(0)} = \text{MLP}_l(\mathbf{x}_l), \;
        \mathbf{h}_p^{(0)} = \text{MLP}_p(\mathbf{x}_p)
    \end{equation*}
    where $\text{MLP}$ indicates a 2 layer multi-layer perceptron network.
    
In the \textbf{message passing} phase, Gated Recurrent Units (GRU) are used to update the path and link state embeddings. 
The state of a path is updated using its GRU based on the states of the links it passes through, which is then aggregated to obtain a message from the paths. A similar approach is used to update the states of the paths using the link states.

This cyclic message-passing phase is repeated until convergence or for a predefined number of iterations. Mathematically, in each iteration $t$, the network computes for all $p$ and $l$
    \begin{align*}
        \mathbf{h}_p^{(t)} &= \text{GRU}_p(\mathbf{h}_p^{(t-1)}, \text{agg}_l(\mathbf{h}_l^{(t-1)})) \\
        \mathbf{h}_l^{(t)} &= \text{GRU}_l(\mathbf{h}_l^{(t-1)}, \text{agg}_p(\mathbf{h}_p^{(t)}))
    \end{align*}
    
Finally, in the \textbf{readout} phase, a two-layer MLP is used to estimate the network metrics using the final path states. For the delays, the MLP estimates the effective queue occupancy on each device in the path. This is divided by the corresponding link capacity to get the transmission delays. For, the jitter and packets dropped, the estimation is made directly using the hidden state of the last node in the path. Mathematically, the prediction becomes
    \(
    \hat{\mathbf{y}} = \text{MLP}(\mathbf{H}_p),
    \)
where $\mathbf{H}_p$ aggregates the states of all links in the path $p$ of the flow whose performance is to be predicted.
Due to the lack of correlation of the network metrics, instead of using a single backbone that estimates all metrics, our experiments show that it is more accurate to use separate RouteNet-Fermi models for learning the backbone model for each metric.

\begin{figure}[t]
    \centering
    \begin{subfigure}{0.24\textwidth}
    \includegraphics[width = \textwidth,trim={0 0 0 1cm},clip]{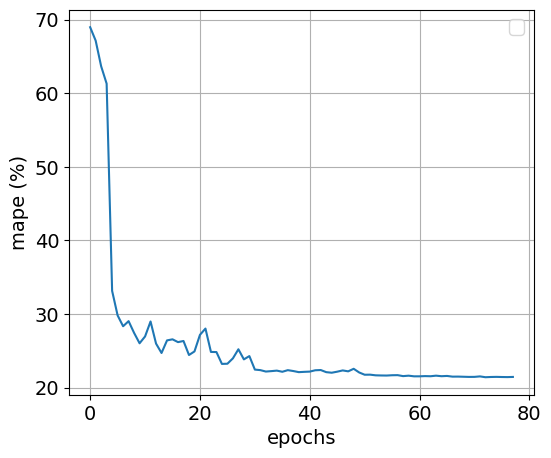} 
    \caption{Delay MAPE converges relatively fast to around 20\%.}
    \label{fig:delay_mape}
    \end{subfigure}
    \begin{subfigure}{0.24\textwidth}
    \centering
    \includegraphics[width = \textwidth,trim={0 0 0 1cm},clip]{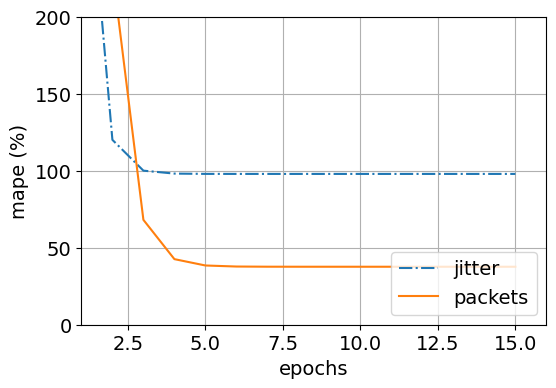} 
    \caption{MAPE for jitter and packets dropped stays high and does not decrease after the first few epochs.}
    \label{fig:pkts_jitter_baseline}
    \end{subfigure}
    \caption{State-of-the-art model RouteNet-Fermi predicts delay well, but not jitter or packets dropped. Many flows experience no jitter or packet drops in our dataset.}
    \label{fig:baselines}
\end{figure}

\subsection{Multi-Metric Estimation}\label{sec:architecture-multi}

We train the RouteNet-Fermi model described above on a real network dataset (see Section~\ref{sec:dataset} for details) in predicting delay, jitter, and packets dropped.  Figure~\ref{fig:delay_mape} shows that this model has fairly low MAPE (mean absolute percent error) when predicting delay, as would be expected since RouteNet-Fermi is designed to do so. However, it performs poorly when predicting jitter and packets dropped, as seen in Figure~\ref{fig:pkts_jitter_baseline}. Indeed, the training levels off within the first few epochs.

\begin{figure*}[t]
    \centering
    \includegraphics[width=0.9\linewidth]{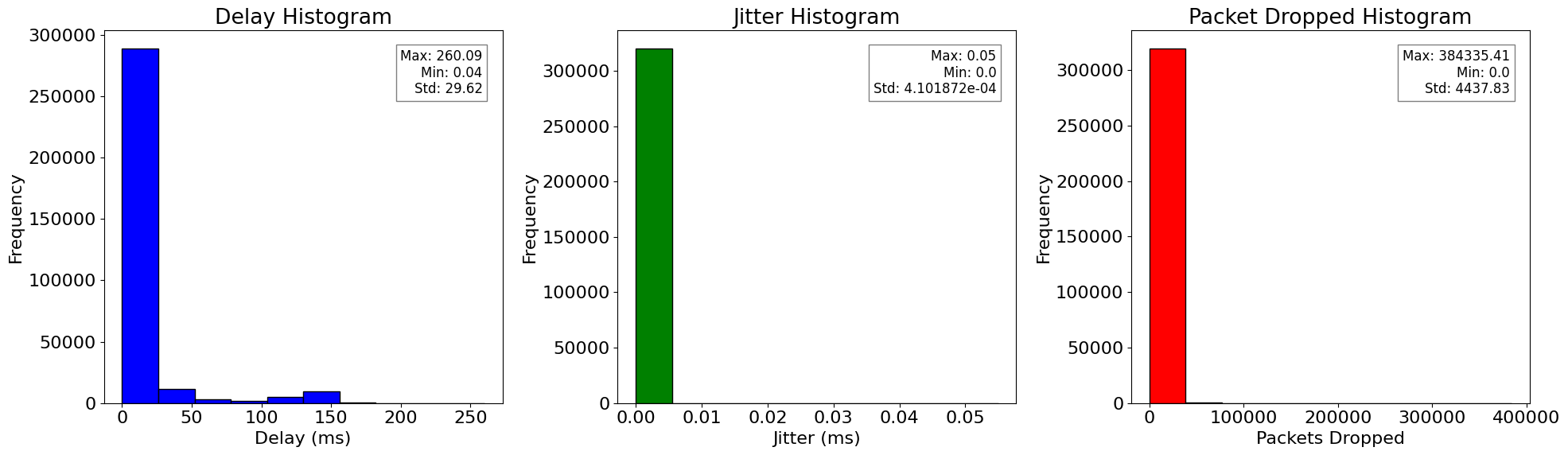}
    \caption{Histogram of delay, jitter and packets dropped in the training data. We see that all metrics are concentrated around zero, in particular the jitter and packets dropped. This concentration around zero makes it challenging to train predictors for these metrics, motivating Figure~\ref{fig:hierarchical_arch}'s hierarchical approach.}
    \label{fig:histogram_metrics}
\end{figure*}

Upon further investigation, we notice that this poor performance may be due to the fact that \textit{most of the flows experience no jitter or packets dropped} (Figure~\ref{fig:histogram_metrics}). We conjecture that this is because of the small and controlled setting of the network testbed. Due to this characteristic of the training data, the RouteNet-Fermi model does not learn useful representations, but simply predicts the same values of 0 for all inputs, resulting in a relatively high MAPE for those flows that experience nonzero jitter or dropped packets. 

\begin{figure}[t]
    \centering
    \includegraphics[width = 0.8\linewidth]{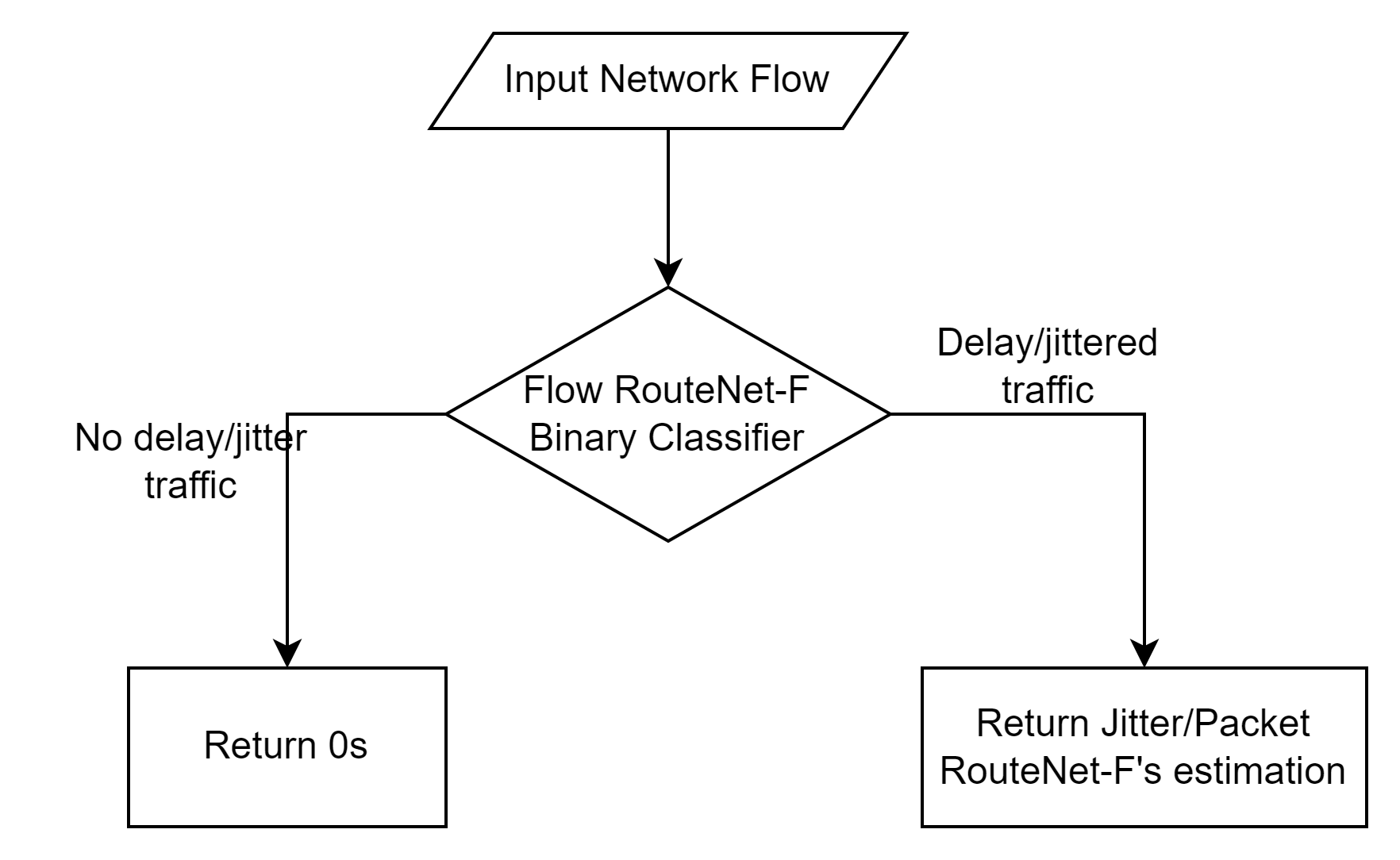}
    \caption{Hierarchical architecture for Jitter/Delay Estimation}
    \label{fig:hierarchical_arch}
\end{figure}

\begin{figure}[t]
    \centering
    \includegraphics[width=0.38\textwidth,trim={0.3cm 0 1cm 1cm}]{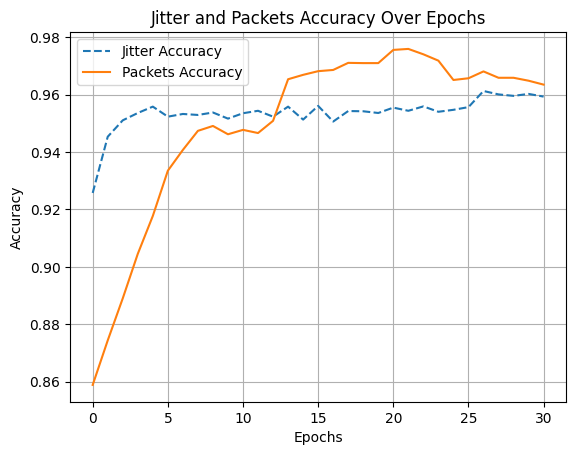}
    \caption{The MLP used in our hierarchical approach can predict whether a flow has zero jitter or packets dropped with high accuracy. We then directly predict 0 values for these flows.}
    \label{fig:jitter_class}
\end{figure}

To prevent the model from making such trivial predictions, we propose a \textit{hierarchical} approach in M3Net where we first train a binary classifier to determine if the traffic would have zero jitter/packets dropped or not. For flows classified as having zero jitter/packets dropped, we simply return 0; otherwise, we use the RouteNet Fermi model to estimate the metrics. This hierarchical setup is illustrated in Figure \ref{fig:hierarchical_arch}. In Figure~\ref{fig:jitter_class}, we show that our classifier can predict whether a flow has zero jitter or packets dropped with more than 95\% accuracy, indicating that this simple addition to the GNN architecture can significantly improve prediction accuracy. 


\subsection{M3Net Architecture}\label{sec:architecture-gated}

We extend the hierarchical architecture that we introduce in the previous section to a mixture-of-experts-based model that can use \textit{multiple} models to predict network performance in different network conditions. Our approach in Section~\ref{sec:architecture-multi} can be thought of as using two experts, one of them trivial, to predict jitter and packets dropped; however, given the range of possible network environments (e.g., with many different paths through the network, levels of congestion, etc.), we conjecture that more experts may produce better prediction results. 

\textbf{Mixture-of-Experts Intuition.} Our approach to achieving this goal is inspired by that of \cite{shazeer2017outrageously}, which introduced a novel approach to conditional computation in neural networks, aimed at increasing model capacity without a proportional rise in computational cost. The authors address the limitation of traditional neural networks, whose capacity to absorb information is restricted by their number of parameters. The proposed approach, the Sparsely-Gated Mixture-of-Experts (MoE) layer, consists of up to thousands of feed-forward sub-networks or "experts," with a trainable gating network that determines a sparse combination of these experts for each example. This method allows for significant improvements in model capacity, achieving over 1000x capacity enhancements while maintaining computational efficiency on modern GPU clusters. The MoE layer has been particularly effective in language modeling and machine translation tasks, where model capacity is crucial for handling large datasets. We therefore expect that it will be useful for our prediction tasks as well.

\textbf{Experts in M3Net.} For a M3Net model with $n$ experts, the output \( y \) of the MoE module is expressed as:
\begin{equation}
y = \sum_{i=1}^{n} G(x)_i E_i(x)
\end{equation}
where \( G(x) \) is the output of the gating network and \( E_i(x) \) is the output of the \( i \)-th expert network for a given input \( x \).
The gating network is computed as:
\begin{equation*}
G(x) = \text{Softmax}(\text{KeepTopK}(H(x), k)).
\end{equation*}
Here we ensure sparsity by defining the function
\begin{equation*}
\text{KeepTopK}(v, k)_i = \begin{cases} 
v_i & v_i \in\left\{\text{top } k \text{ elements of } v\right\}, \\
-\infty & \text{otherwise}.
\end{cases}
\end{equation*}
The original MoE paper~\cite{shazeer2017outrageously} used a linear function
$H(x)_i$, so that the weight of each expert is determined by a linear function of the input features. In our architecture, we take $H(x)_i$ to be a MLP, allowing for more expressive relationships between the input features and the gating function. Also, in the original paper Gaussian noise is added to encourage load balancing, but we do not observe such issue across our experts and do not add the noise term.
Intuitively, the final prediction is a weighted sum of the prediction of each expert, $E_i(x)$. Since the coefficients of the gating MLP are learnable parameters, the model effectively learns the different conditions of $x$ under which each expert performs best and should be weighted most.

\begin{figure*}[ht]
    \centering
    \includegraphics[width=\linewidth]{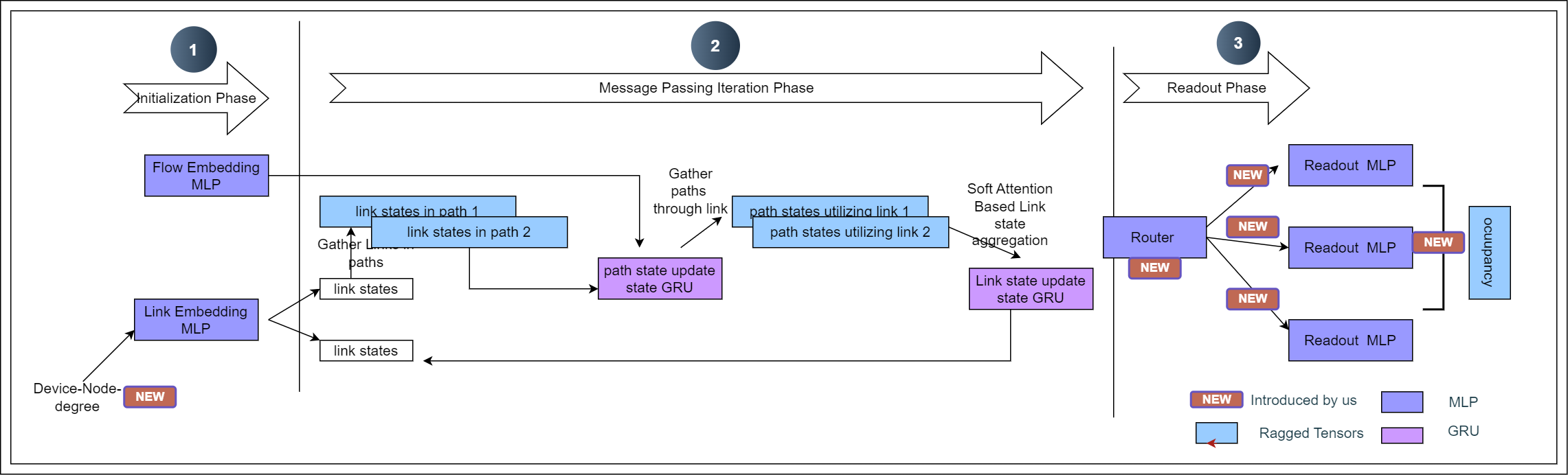}
    \caption{M3Net Architecture. We initialize the network with flow- and link-specific states, both of which are passed through separate MLPs. GRUs are then used for message passing through the links along the flow's path, and the output goes through the gating router to the readout MLPs. These MLPs' outputs are then combined into the final prediction.}
    \label{fig:gatedroutenetarchiecture}
\end{figure*}

We draw a similar intuition and incorporate multiple experts trained on various network traffic scenarios, and the MoE can assess a network's performance under diverse conditions. The gating network dynamically selects the most relevant experts to analyze specific traffic scenarios. Intuitively, we can either choose the experts  \( E_i(x) \) to be a complete RouteNet Fermi model or the Readout MLPs. We decide on the latter option due to its parameter efficiency, scalability and early experiments of small subsets of the data. On the gating network, we use an MLP of the same size as the actual MLP readouts with a softmax activation for assigning probabilities to the different MLP readout networks. This architecture is summarized in the readout phase of Figure \ref{fig:gatedroutenetarchiecture}.

\subsection{Network Training}\label{sec:network training}
We train each model weight in the M3Net architecture, including the weights of the gating MLP, by backpropagating the loss from Equation (\ref{eq:1})'s objective function. Since we use different mixtures of experts for each of our three performance metrics, the weights of the gating MLP and expert MLPs are updated separately for each metric; however, all three share the same model weights for the initialization and message passing phases. 
While any neural network optimizer can be used to train this network, in our experiments, we use an AdamW Optimizer with a learning rate of 0.01. A learning rate scheduler is used to decrease the learning if the validation loss stagnates. The network is trained for 12 message-passing iterations with a hidden dimension of 16 for all the modules. 

%% file: sections/implementation.tex
\section{Efficient GPU Implementation}\label{sec:implementation}

Having detailed our M3Net architecture and training setup, we now turn to efficient implementation of the training process. We identify two main areas of speedup over prior work.

\textbf{GPU implementation.} From our study, all the implementations of the RouteNet architecture only run on CPU and do not achieve any speed up on GPUs~\cite{ferriol2023routenet,modesto2023graph}. This is because the \textit{batch size is not controllable}, but determined by the number of flows used for each network configuration. For each network setup, a number of flows are sent through the network by the data collectors. Current implementations simply use these flows for a given experiment as one batch on which to run gradient descent to train the predictor model.

This na\"ive implementation may lead to bias and slow convergence. Varying the batch size for each round of gradient descent can slow convergence due to inducing greater variance during the training, and sampling all flows within a batch from a single network configuration can also lead to high variance in the gradient direction from iteration to iteration as the network confiuration changes. Even worse, \textit{the variable number of flows per batch precludes a GPU-based implementation}, significantly slowing training times in practice. To the best of our knowledge, however, this flow batching is the dominant paradigm in prior GNN implementations for network prediction, as it simplifies the message passing stage by ensuring there are a constant number of competing flows and paths, as well as a consistent network topology, for each sample (each flow) within a batch.

\begin{figure}[!htbp]
    \centering
    \includegraphics[width=0.49\textwidth]{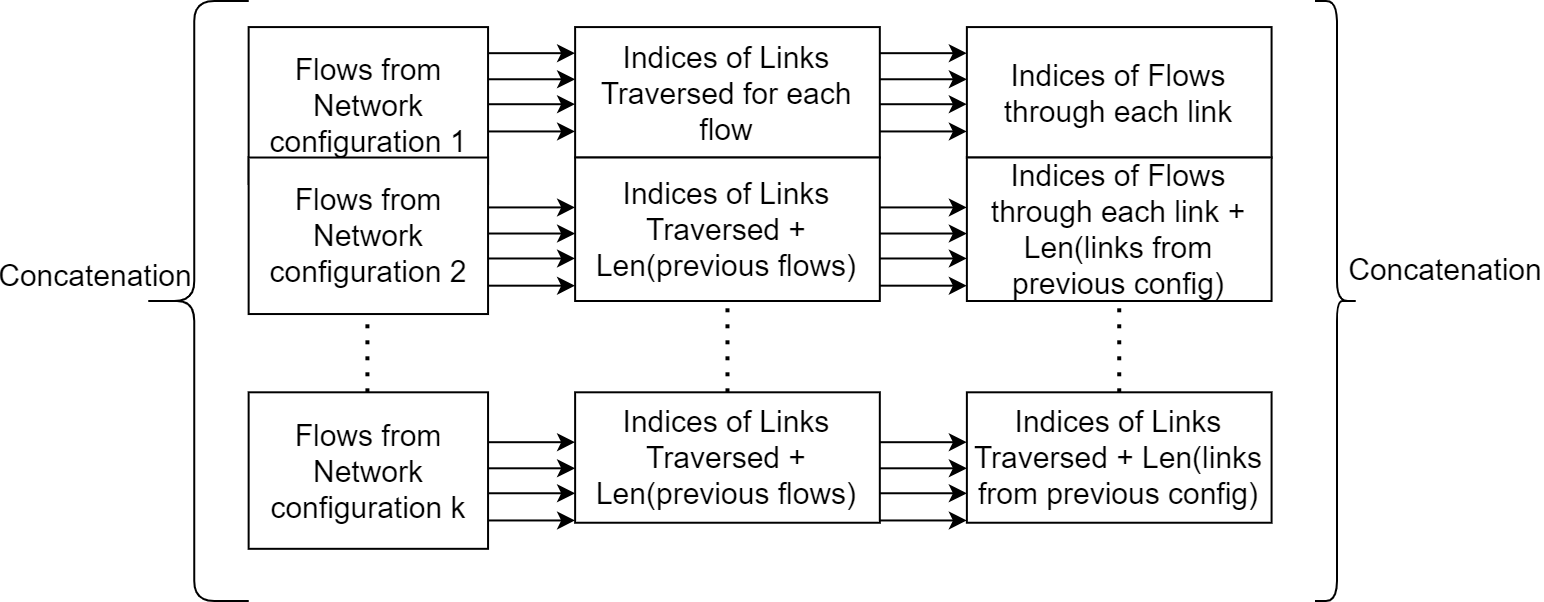}
    \caption{The batch extending approach by concatenation and updating indices, allowing us to sample constant-size batches from multiple experiment configurations.}
    \label{fig:gpu_eff}
\end{figure}

To solve this challenge, we concatenate flows and links from different experiment configurations while expanding and updating the link and flow indices correspondingly to match the new size of the flows and links matrices. In our implementation, we set a parameter to determine the number of flows to concatenate. This is illustrated in Figure \ref{fig:gpu_eff}. With this concatenation, we can choose batches that include a constant number flows from different experiment configurations, facilitating a GPU implementation and eliminating problems of bias and slow convergence.
The speed up for concatenating several number of flows together is shown in Table \ref{efficiency_table}. As can be seen, merging a larger number of flows can dramatically decrease the time for each training epoch.

\begin{table}[htbp]
\caption{Speed up from merging experiment flows.}
\centering
\begin{tabular}{|c|c|c|}
\hline
Flows Merged & Time for Epoch (s) & Average Samples per Batch \\
\hline
1 & 389.33 & 75.74  \\
\hline
2 &188.82 & 151.5 \\
\hline
4 & 167.90 & 302.90 \\
\hline
8 & 151.92 & 605.80  \\
\hline
16 & 145.13 & 1211.61  \\
\hline
\end{tabular}
\label{efficiency_table}
\end{table}

\textbf{Pre-processing feature extractions.} During the initialization phase, na\"ive implementations would require several feature extraction steps that are repeated after each epoch. We identify and pre-compute the following feature steps instead of repeating them during each epoch. Note that these pre-computations \textit{do not change the input feature values}; they simply avoid redundant computations.

\begin{itemize}
    \item \textit{Feature Normalization:} Since the statistics (such as Mean, Max, Standard deviation) for normalization are available from the whole training data, we can pre-normalize all features instead of redoing the normalization for the training data used in each epoch. 
    \item  \textit{Link Load computation:} The load of link $l$ is defined as:
    \begin{equation}
    L_l =  \sum_{i = 1}^{n} F_{i}/ link\_capacity(L) 
    \end{equation}{}

    where $F_{i}$ is the average traffic bandwidth and $n$ is the total number of flows that pass through link $l$. We can pre-compute this quantity for all links, instead of re-computing it for all links along each flow's path, every time we encounter that flow.
\end{itemize}

%% file: sections/dataset.tex
\section{Dataset Description}\label{sec:dataset}
\begin{figure}
    \centering
    \includegraphics[width=0.7\linewidth]{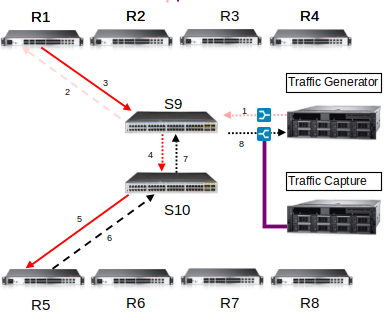}
    \caption{Interconnection of devices in the testbed on which our training data is generated~\cite{gnn-challenge}.}
    \label{fig:arch}
\end{figure}

We use a dataset from~\cite{gnn-challenge} to evaluate M3Net. The \textbf{testbed network} on which the dataset is generated is made up of 2 Huawei CloudEngine S5732 switches and 8 Huawei Net Engine 8000 M1A routers, traffic generator\cite{trex} and a traffic capture device. Each of the routers has six ports and each switch has 48 ports. Each router is connected to a neighbouring router through a switch, with a maximum node degree of 5 because one of the six ports is reserved for traffic generation. The devices are interconnected as shown in Figure \ref{fig:arch}.  The dataset was generated from 11 random topologies of the devices across 10 different routing configurations. Each topology is made up of 5 to 8 nodes. During each experiment, traffic is generated at the traffic generator, sent to the source node, and routed to the destination node (which is always the traffic generator). The traffic capture device captures the packet traces of the traffic which is used to record 18 performance metrics relating to the delay, jitter and the number of packets dropped. This is illustrated in Figure 2 below with traffic flowing from the Traffic Generator (TG) to R1-S9-S10-R5-S10-S9-TG.  Each experiment is 10 seconds long, however, only the last 5 seconds were recorded because the network is in a transient state in the first five seconds. 

During \textbf{traffic generation}, the packets are either sent Multi-burst mode (MB) or a mixture of both MB and Constant Bit Rate (CBR). In the constant bit rate, traffic is sent in short constant time intervals while in multi-burst mode packets are sent in bursts defined by an inter-burst gap time and the number of packets per burst. An example of this MB traffic is on-demand streaming, where buffering can occur, traffic might be sent in bursts to fill buffers efficiently. However, during each flow, the packet size is constant in both modes. For each configuration of the network (a specific topology and routing configuration), several sets of flows of different rates are sent and the corresponding performance metrics are recorded. The total size of the dataset generated is 500GB across 4388 unique network configurations.

%% file: sections/evaluation.tex
\section{Evaluation}\label{sec:evaluation}

We finally evaluate our proposed M3Net architecture on Section~\ref{sec:dataset}'s dataset. We first demonstrate our ability to predict multiple performance metrics in Section~\ref{sec:eval-multi} and then show that M3Net outperforms the predictions of prior works.

\subsection{Multi-Metric Prediction Accuracy}\label{sec:eval-multi}
\begin{figure}[t]
    \centering \includegraphics[height=0.2\textheight]{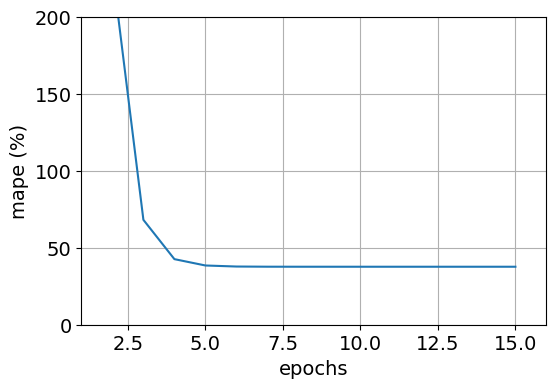} 
    \caption{MAPE of the packets dropped, using Figure~\ref{fig:hierarchical_arch}'s hierarchical model to filter out flows with no packets dropped. We observe a high MAPE of 31\%.}
    \label{fig:pkts_hierarchical}
\end{figure}

\begin{figure}[t]
    \centering
    \includegraphics[width=0.35\textwidth]{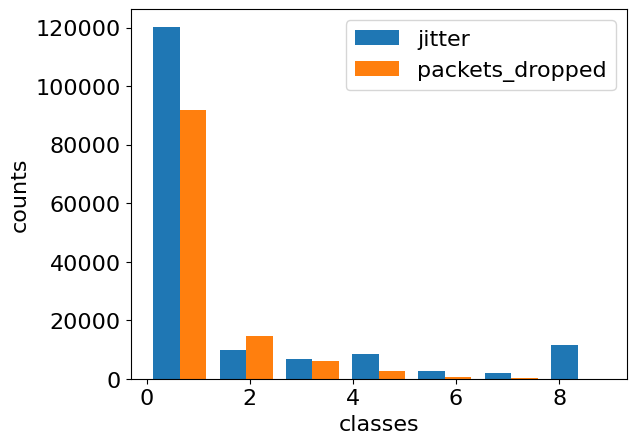}
    \caption{Distribution of the fraction of packets dropped and jitter, binned into a granularity of 0.1, omitting flows with 0 jitter or packets dropped. We convert the problem of predicting the fraction of packets dropped to that of classifying whether a flow's fraction of packets dropped lies in each bin.}
    \label{fig:pkts_distributions}
\end{figure}

A straightforward prediction of jitter and packets dropped would treat both as regression problems. However, we find that we obtain higher accuracy when treating them as classification problems. Figure~\ref{fig:pkts_hierarchical} shows the MAPE when we use Figure~\ref{fig:hierarchical_arch}'s hierarchical approach and a regression MLP in the readout phase to predict the number of packets dropped for a flow. We see that we obtain a high MAPE of 31\%, though this result is better than the 51\% MAPE achieved without the hierarchical filter, shown in Figure~\ref{fig:pkts_jitter_baseline}. Thus, we can expect to make more accurate predictions if we bin the fraction of packets dropped into classes of granularity 0.1 and classify each flow into one of these bins. Correct classification would then correspond to a MAPE of at most 10\%, considerably lower than the 31\% obtained with a regression model.
The distribution of the resulting classes is shown in Figure \ref{fig:pkts_distributions}. We find similar results for jitter, with a 47\% MAPE using a regression readout MLP, and thus follow the same binning and classification approach, with class distribution shown in Figure \ref{fig:pkts_distributions}.


\begin{figure*}[t]
\centering
\begin{subfigure}{0.32\textwidth}
    \includegraphics[width=0.99\textwidth,trim={0.3cm 0 1cm 1cm},clip]{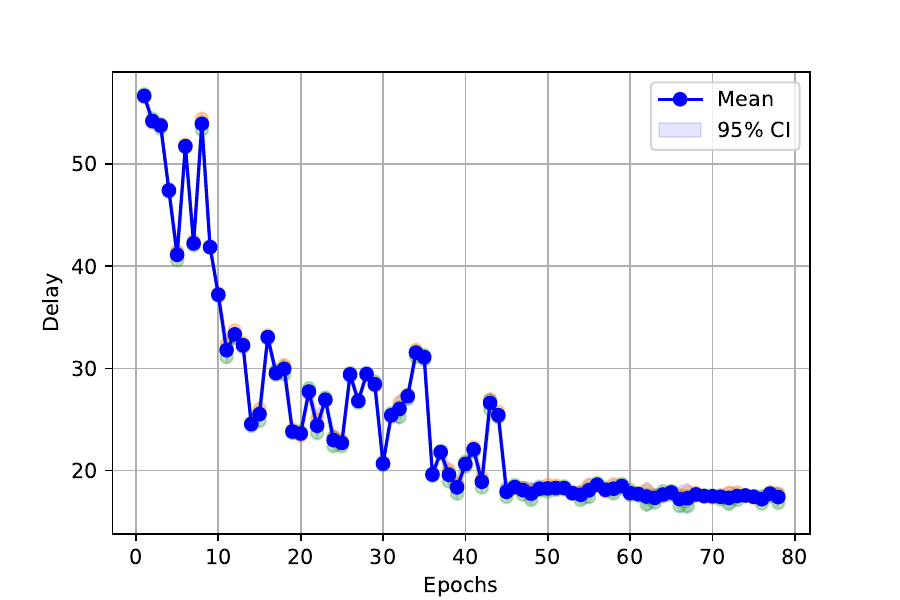}
    \caption{MAPE of delay predictions.}
    \label{fig:delay_confidence}
\end{subfigure}
\begin{subfigure}{0.32\textwidth}
    \includegraphics[width=0.99\textwidth,trim={0.3cm 0 1cm 1cm},clip]{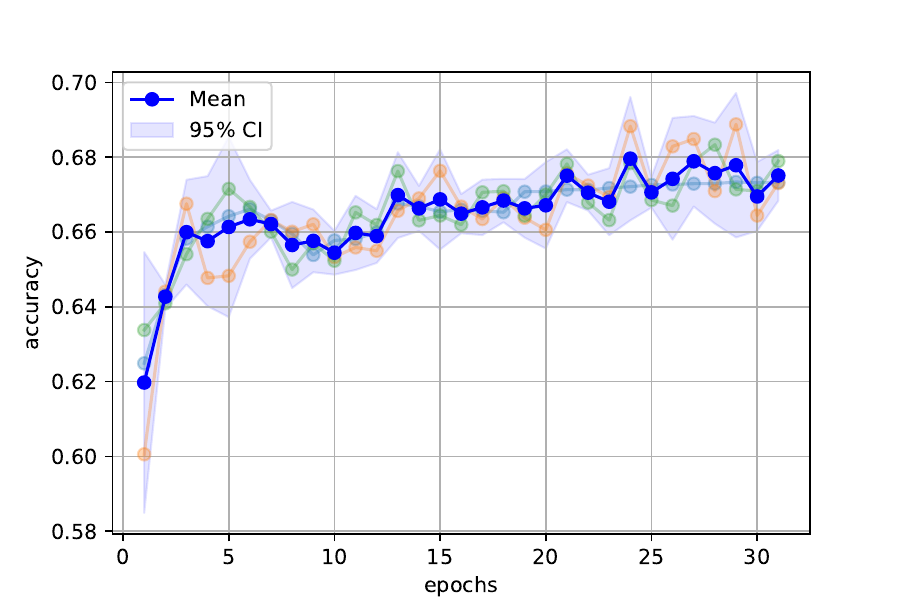}
    \caption{Accuracy of the jitter.}
    \label{fig:jitter_confidence}
\end{subfigure}
\begin{subfigure}{0.33\textwidth}
    \includegraphics[width=0.99\textwidth,trim={0.3cm 0 1cm 1cm},clip]{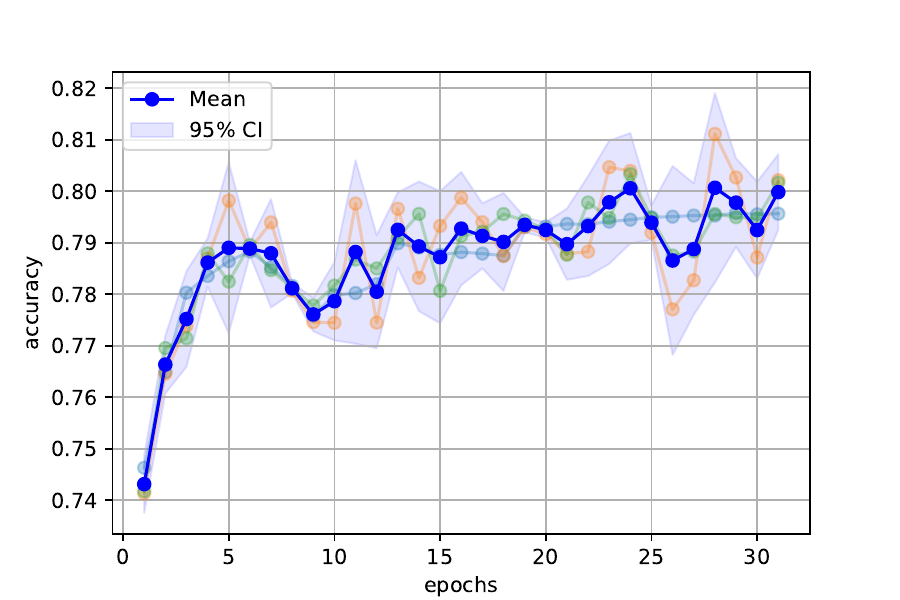}
    \caption{Accuracy of fraction of packets dropped.}
    \label{fig:pkts_confidence}
\end{subfigure}
\caption{Prediction results of our architecture, averaged over three experimental runs. M3Net consistently achieves a MAPE below 20\% for average delay, with accuracies of 78.7\% and 66.47\% respectively for the jitter and packets dropped.}
\end{figure*}

After redefining our predictions of jitter and packets dropped as classification rather than regression problems, we \textbf{evaluate M3Net's success} in predicting average flow delay, jitter, and packets dropped. We use the training dataset in Section~\ref{sec:dataset}. The test data comes from 300 different network configurations, with the training procedure described in Section~\ref{sec:network training}, and four experts. All three experimental runs obtain nearly the same MAPE for delay prediction (Figure~\ref{fig:delay_confidence}). M3Net achieves relatively high 66.47\% accuracy for jitter (Figure \ref{fig:jitter_confidence}) and 78.7\% accuracy (Figure \ref{fig:pkts_confidence}) in this setting in classifying packets dropped.

\begin{figure}[t]
    \centering \includegraphics[width=0.3\textwidth,trim={0.3cm 0 1cm 1cm}]{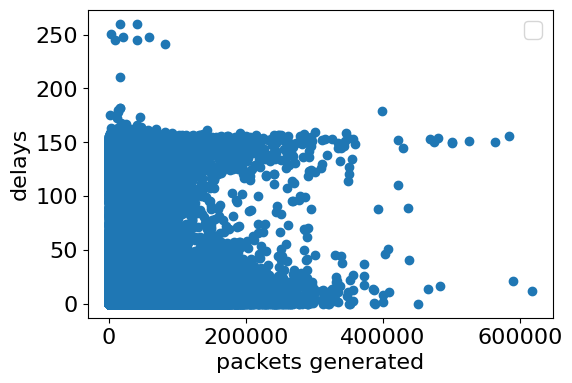}
    \caption{Each point represents one flow. We observe no correlation between numbers of packets generated and delays.}
    \label{fig:pktsvsdelay}
\end{figure}

\begin{figure}[t]
    \centering \includegraphics[width=0.35\textwidth,trim={0.3cm 0 1cm 0.8cm}]{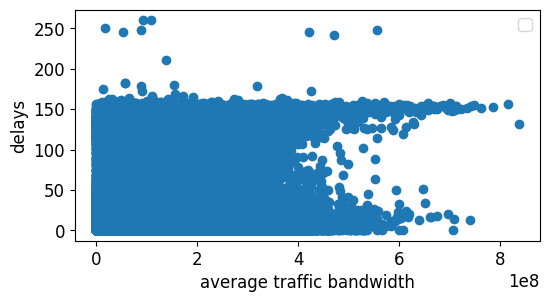}
    \caption{Each point represents one flow. We observe no correlation between average traffic bandwidth and delays.}
    \label{fig:band}
\end{figure}

We can further use M3Net to \textbf{explore the relationships} between different types of flows and their performance. While we might intuitively expect that larger flows with more packets generated would be correlated with longer delays, due to higher congestion, Figure~\ref{fig:pktsvsdelay} shows that this is not the case: we observe no correlation. Similarly, Figure~\ref{fig:band} shows that there is no visible correlation between the average traffic bandwidth and flow delay, even though we might expect a higher bandwidth to correlate with lower delays due to the network having more capacity. 

\subsection{Comparison to Baselines}

We next compare M3Net to three baselines: a simple MLP, RouteNet-Fermi~\cite{ferriol2023routenet}, and RouteNet-Fermi with attention~\cite{dhamala2024performance}. The latter two are state-of-the-art models for predicting average flow delay, and as described in Section~\ref{sec:architecture} we build on their architecture. Since RouteNet-Fermi performs poorly in predicting jitter and packets dropped (Figure~\ref{fig:pkts_jitter_baseline}), we focus on comparing these algorithms based on the average delay. 
To evaluate the M3Net's MoE approach, we also experiment with having 4 or 8 experts. 

\begin{figure}[t]
    \centering \includegraphics[width=0.46\textwidth,trim={0.3cm 0 1cm 1cm},clip]{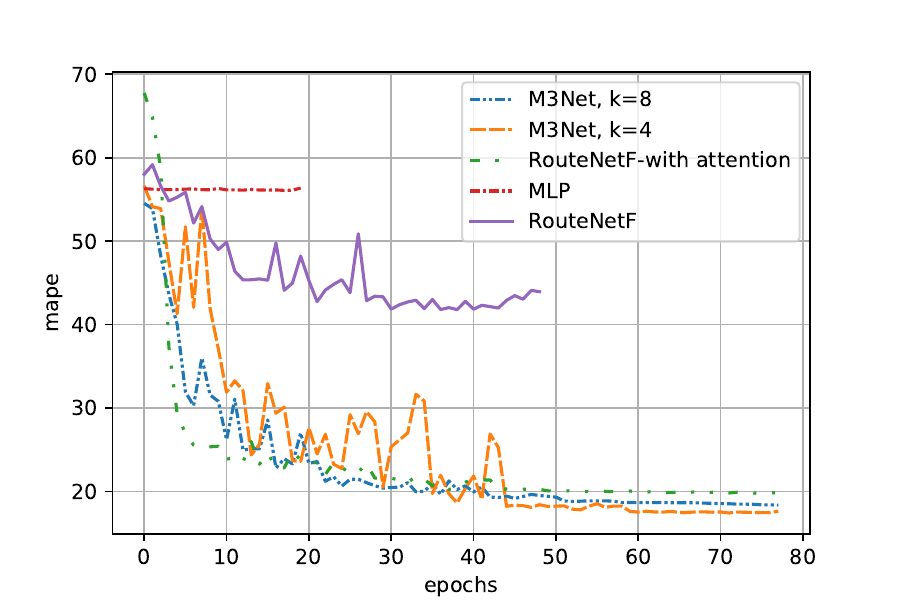}
    \caption{Results of the M3Net architecture with 4 and 8 experts}
    \label{fig:delay_gated}
\end{figure}

Figure~\ref{fig:delay_gated} shows that our \textbf{M3Net  architecture outperforms all other baselines}. At the start (0-10 epochs), all models show a high validation loss, which gradually decreases. The RouteF-with attention model initially performs better than both M3Net models, but this changes as the training progresses. Between 10 to 40 epochs, the M3Net models show more variability in test loss compared to RouteF-with attention. This suggests that the gating mechanism and the number of experts being dynamically selected are influencing the stability of the learning process. Interestingly, M3Net with 4 experts outperforms the model with 8 experts, despite the latter having more capacity to learn. This suggests that increasing the number of experts does not necessarily lead to better performance. The additional complexity from more experts may introduce noise or overfitting. The superior performance of the 4-expert model underscores the importance of finding the right balance to mitigate overfitting.

The training times shown in Table \ref{tab:performance} show that M3Net does not induce much increase training time over the base RouteNet Fermi with attention model. The MLP model has a much faster training time because it does not take into account any of the graph structure or link features, but it has correspondingly lower delay MAPE. The average delay MAPE over 3 runs is shown in Table \ref{tab:performance}.  Overall, our results indicate that M3Net, with an appropriately chosen number of experts, can significantly enhance network performance assessment under varying traffic conditions, validating our approach and choice of model components. 


\begin{table}[h!]
\caption{M3Net has lower MAPE in predicting delay (averaged over 3 runs) but slightly longer training time.}
    \centering
    \begin{tabular}{|c|c|c|}
        \hline
        \textbf{Method} & \textbf{Delay MAPE} & \textbf{Average Training Time/epoch (s)} \\
        \hline
        M3Net & 17.39 & 389.33 \\ 
        \hline
        RouteNet F/Attn & 20.06 & 356.22 \\
        \hline
        MLP & 56.61 & 10.33 \\ 
        \hline
        RouteNet F & 43.38 & 291.66 \\ 
        \hline 
    \end{tabular}
    
    \label{tab:performance}
\end{table}